\documentclass[12pt]{article}
\usepackage{epsfig}
\newcommand{\be}{\begin{equation}}
\newcommand{\ee}{\end{equation}}
\newcommand{\bea}{\begin{eqnarray}}
\newcommand{\eea}{\end{eqnarray}}

\title{Sidestepping the Cosmological Constant\\
with Football-Shaped Extra Dimensions}

\author{Sean M. Carroll\thanks{\tt carroll@theory.uchicago.edu} 
\ and Monica M. Guica\thanks{\tt mmguica@uchicago.edu} \\ \\
\it Enrico Fermi Institute, Department of Physics, \\
\it and Center for Cosmological Physics, University of Chicago \\
\it 5640 S.~Ellis Avenue, Chicago, IL~60637, USA \\ \\
hep-th/0302067. EFI-2003-05}
\begin{document}
\baselineskip 16pt
\maketitle

\begin{abstract}
We present an exact solution for a factorizable brane-world spacetime
with two extra dimensions and explicit brane sources.  The
compactification manifold has the topology of a two-sphere, and is
stabilized by a bulk cosmological constant and magnetic flux.  The
geometry of the sphere is locally round except for conical
singularities at the locations of two antipodal branes, deforming the
sphere into an American-style football.  The bulk magnetic flux needs
to be fine-tuned to obtain flat geometry on the branes.  Once this is
done, the brane geometry is insensitive to the brane vacuum energy,
which only affects the conical deficit angle of the extra dimensions.
Solutions of this form provide a new arena in which to explore
brane-world phenomenology and the effects of extra dimensions on the
cosmological constant problem.
\end{abstract}

\vfill\eject

The idea that Standard Model fields are confined to a brane 
embedded in (possibly large) extra dimensions has opened up
new possibilities in particle physics, gravitation, and cosmology
\cite{rv,hw,add,rs,review}.  In particular, novel interactions between
matter and spacetime curvature in brane-world models enable
new attacks on longstanding puzzles such as 
the hierarchy problem and cosmological constant problem.

In this paper we present an exact solution to Einstein's equation
representing two three-branes at antipodal points of two extra
dimensions with spherical topology.  The branes are modeled as
uncharged delta-function energy-momentum distributions.
The geometry factorizes
into flat Minkowski spacetime on the branes and a football-shaped
compactification manifold (a round sphere with a wedge removed and
opposite sides identified).  The compact dimensions are stabilized
by a competition between a bulk cosmological constant, magnetic flux,
and spatial curvature.

Our initial interest in spacetimes of this sort was motivated by the
desire to find simple exact solutions, including stabilizing bulk
fields and explicit brane sources, representing factorizable brane-worlds
of the ADD type (named after the authors of the first paper in
\cite{add}).  The solution we find has a 
remarkable property:  the brane geometry is insensitive to the
value of the brane tension (what a four-dimensional 
observer would calculate as the vacuum energy).  Instead, the only
effect of the tension is to change the deficit angle associated with
the branes; the bulk geometry alters its shape to absorb 
the brane cosmological constant.
This phenomenon does not by itself constitute a solution to the 
cosmological constant problem, which in this context manifests itself
as the need to finely-tune magnetic flux in terms of the bulk
vacuum energy.  Nevertheless, the ability to push 
effects of the cosmological constant
off into the extra dimensions changes the nature of the problem in
an interesting way, suggesting the possibility that the vacuum 
energy is not actually small, but simply invisible to four-dimensional
observers.

We begin by considering stabilization of homogeneous extra dimensions,
not yet including brane sources, a subject which has been
extensively studied \cite{hpcs,freundrubin,rss,gz,sundrum,cghw}.  
Since our motivation comes from
``large'' extra dimensions, we work in classical general relativity,
imposing the weak energy condition (WEP) on bulk fields.  (Our
ultimate conclusions will rely on classical
general relativity, but not on the size of the extra dimensions.) 
We consider a factorizable geometry $M\times\Sigma$, where $M$
is the macroscopic $(3+1)$-dimensional universe, and $\Sigma$ is
a two-dimensional compactification manifold.  Upon compactification,
the scale factor of the extra dimensions becomes a scalar field,
the dilaton.  With classical
fields obeying the WEP, insisting that the dilaton have
positive (mass)$^2$ requires
that $\Sigma$ have a net positive curvature \cite{cghw}.

A specific mechanism for stabilization invokes a cosmological
constant and a magnetic flux in the bulk.
The bulk action can be written
\be
  S_6 = \int d^6X\sqrt{|G|}\, \left( {1\over 2}M_6^4 R - \lambda
  - {1\over 4} F_{ab} F^{ab}\right)\ ,
  \label{s6}
\ee
where the $X^a$'s, $G_{ab}$, $M_6$, and $\lambda $ are
the six-dimensional coordinates, metric, 
reduced Planck mass, and vacuum energy 
density, respectively.  We decompose the coordinates into 
the four macroscopic dimensions $x^\mu$ and the two extra dimensions
$y^i$, and consider an ansatz for which the geometry factorizes
into a flat metric $\eta_{\mu\nu}$ on $M$ and a metric 
$\gamma_{ij}$ on $\Sigma$,
\be
  ds^2 = G_{ab}\, dX^a dX^b = \eta_{\mu\nu}\, dx^\mu dx^\nu
  +\gamma_{ij}(y) dy^i dy^j\ .
  \label{metric}
\ee
We furthermore take the gauge field to consist of magnetic
flux threading the extra-dimensional space, 
so that the field strength takes the form
\be
  F_{ij} = \sqrt{\gamma} B_0 \epsilon_{ij}\ ,
\ee
where $B_0$ is a constant, $\gamma$ is the
determinant of $\gamma_{ij}$, and the antisymmetric symbol is
normalized to $\epsilon_{12}=1$; all other
components of $F_{ab}$ vanish.  (In the quantum theory, there will
be a quantization condition on $B_0$ if the gauge group is compact;
we do not worry about this issue in this paper.)
It is straightforward to verify that
this ansatz satisfies Maxwell's equations,
\be
  \nabla_a F^{ab} = 0\ ,\qquad \nabla_{[a}F_{bc]}=0\ .
\ee
The other equation we have to satisfy is Einstein's equation,
which may be written 
\be
  R_{ab} - {1\over 2} Rg_{ab} = M^4_6 T_{ab}\ .
  \label{einstein}
\ee 
The energy-momentum tensor is a sum of contributions from
the bulk cosmological constant and the gauge field,
\be
  T_{ab} = T_{ab}^{\lambda} + T_{ab}^{F}\ ,
\ee
for which the explicit forms are
\bea
  T_{ab}^{\lambda} &=& -\lambda\left(\matrix{ \eta_{\mu\nu} & 0 \cr
  0 & \gamma_{ij} }\right)\cr
  T_{ab}^{F} &=& -{1\over 2} B_0^2\left(\matrix{\eta_{\mu\nu} & 0 \cr
  0 & -\gamma_{ij} }\right)\ .
\eea

A static, stable solution \cite{gz,cghw} is obtained by choosing the
extra-dimensional space to be a two-sphere,
\be
  \gamma_{ij}(y) dy^i dy^j = a_0^2(d\theta^2 + \sin^2\theta \,
  d\phi^2)\ ,
\ee
and fixing the magnetic field strength $B_0$ and the radius $a_0$
in terms of the bulk cosmological constant,
\be
  B_0^2 = 2\lambda\ ,\qquad
  a_0^2 = {M_6^4 \over 2\lambda}\ .
  \label{abtune}
\ee
The radius $a_0$ will dynamically adjust itself to the static
solution; the magnetic flux, however, requires an unavoidable fine-tuning.
A different value of $B_0$ would induce a de~Sitter or 
anti-de~Sitter geometry in the large dimensions; this tuning
therefore simply reflects the cosmological constant problem.

We now add branes to this solution.  The brane action
can be written either as a four-dimensional integral over
an ordinary Lagrange density $\widehat{\cal L}_4$, or as a six-dimensional
integral over a distributional density $\widetilde{\cal L}_4$,
\be
  S_4 = \int d^4x\sqrt{|g|}\, \widehat{\cal L}_4 
  = \int d^6X\sqrt{|G|}\, \widetilde{\cal L}_4 \ ,
\ee
where $g_{\mu\nu}$ is the metric pulled back to the brane worldvolume.
The regular and distributional actions can be related by an integral
over the brane worldvolume,
\be
  \widetilde{\cal L}_4 = \int d^4x \sqrt{|g|\over |G|}\, 
  \widehat{\cal L}_4 \delta^{(6)}(X-X(x))\ .
\ee
In this paper we will only consider the tension of the branes, not
any localized matter fields, in which case we have
\be
  \widehat{\cal L}_4 = -\sigma\ ,
\ee
where $\sigma$ is the brane tension.  For a collection of parallel,
equal-tension branes with Minkowski symmetries along their 
worldvolumes, the brane energy-momentum tensor in the geometry
(\ref{metric}) takes the form
\be
  T_{ab}^{\rm branes} = -{\sigma\over\sqrt{\gamma}}
  \left(\matrix{\eta_{\mu\nu}  & 0 \cr 0 & 0 }\right)
  \sum_n\delta^2(y_n)\ ,
\ee
where the branes are located at positions $y_n$.

The simplest way to introduce branes into the stabilized geometry
just considered is to place two branes at opposite
poles of the spherical extra dimensions. 
Although the extra dimensions are compact, it is convenient to 
represent them as a conformal factor times a flat plane in polar
coordinates,
\be
  \gamma_{ij} dy^i dy^j = \psi(r)(dr^2 + r^2 d\phi^2)\ .
\ee
One brane is at $r=0$, the other at $r=\infty$ (or more properly,
we require a new coordinate patch to cover the south pole).  Despite the
apparent asymmetry, for an appropriate conformal factor $\psi(r)$
the geometry will be invariant under reflections about the
equator of the two-sphere.  Note that
the two-dimensional delta-function (which enters the brane
energy-momentum tensor) is conveniently represented as
\be
  \delta^2(y) = {1\over 2\pi}\nabla^2\ln{r}\ .
\ee
Here, $\nabla^2f = f'' + {1\over r}f'$,
where a prime indicates differentiation with respect to $r$ (we
assume the angular derivative vanishes due to rotational symmetry
in the $\phi$ direction). 

For this metric,
% we have the following nonvanishing connection
%coefficients,
%\bea
%  \Gamma^r_{rr} &=& {1\over 2}{\psi' \over \psi}\cr
%  \Gamma^r_{rr} &=& -r\left(r{\psi' \over 2\psi} +1\right)\cr
%  \Gamma^r_{rr} &=& {1\over r}\left(r{\psi' \over 2\psi} +1\right)\ ,
%  \label
%\eea
the Ricci tensor and scalar are
\bea
  R_{\mu\nu} &=& 0 \cr
  R_{rr} &=& -{1\over 2} \nabla^2 \ln{\psi}\cr
  R_{\phi\phi} &=& -{r^2\over 2} \nabla^2 \ln{\psi}\cr
  R &=& -{1\over \psi} \nabla^2 \ln{\psi}\ .
  \label{ricci}
\eea
Einstein's equation, given by (\ref{einstein}), has only
two independent pieces:  from the $\mu\nu$
(longitudinal) components, and from the $ij$ (transverse) components.
The longitudinal equation is
\be
  {M_6^4\over 2\psi} \nabla^2\ln{\psi} = -\lambda - {1\over 2}B_0^2
  - {\sigma \over 2\pi \psi}\nabla^2 \ln{r}\ ,
  \label{e1}
\ee
and the transverse equation gives
\be
  0 = -\lambda + {1\over 2}B_0^2\ .
  \label{e2}
\ee
This expresses the condition that the magnetic field be chosen to 
balance the bulk cosmological constant, exactly
as in (\ref{abtune}).  We choose to use $\lambda$ as our
independent variable, and plug back into (\ref{e1}) to obtain
\be
  M_6^4 \nabla^2 \ln{\psi} = -4\lambda\psi - {\sigma\over \pi}
  \nabla^2\ln{r}\ .
  \label{psieq}
\ee
Comparing this to the expression for the scalar curvature in
(\ref{ricci}), we see that $R$ will be a constant over the extra
dimensions, except at $r=0$ (the locations of the
brane).  Thus, the local geometry of the extra dimensions will 
be perfectly spherical away from the branes, although the global
geometry will be different.

Equation (\ref{psieq}) has previously been considered by Deser and
Jackiw in the context of point-particle solutions to (2+1)-dimensional
gravity with a positive cosmological constant \cite{dj}.
They find the solution
\be
  \psi(r) = {4\alpha^2 a_0^2\over 
  r^2[(r/r_0)^\alpha + (r/r_0)^{-\alpha}]^2}\ ,
\ee
where $r_0$ is an arbitrary parameter characterizing our choice
of coordinates, and $\alpha$ and $a_0$ are observable quantities
given by
\be
  \alpha = 1-{\sigma\over 2\pi M_6^4}\ ,\qquad
  a_0^2 = {M_6^4 \over 2\lambda}\ .
  \label{tune3}
\ee
We see that $a_0$ is the radius of curvature, with the same value
as in (\ref{abtune}).  The parameter $\alpha$ represents the effect
of the branes.  This effect is equivalent to removing from the sphere
a wedge stretching from the north pole to the south pole, and 
identifying opposite sides; this yields the ``football'' geometry
alluded to in our title, and portrayed in Figure One.  
\begin{figure}
  \begin{center}
 \includegraphics[scale=1]{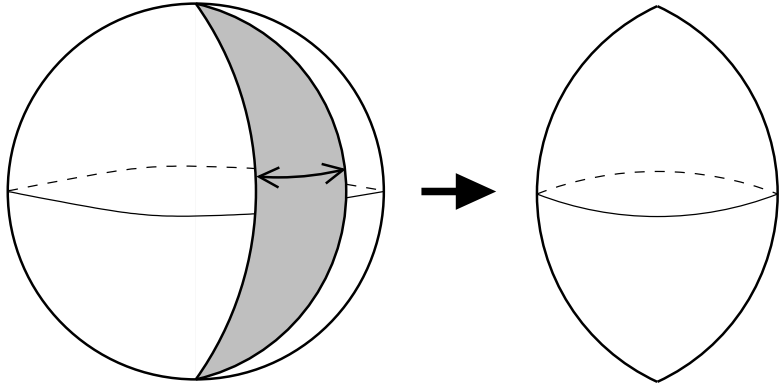}
  \end{center}
  \caption{Removing a wedge from a sphere and identifying opposite
  sides to obtain a football geometry.  Two equal-tension branes with
  conical deficit angles are located at either pole; outside the
  branes there is constant spherical curvature.}
  \label{ecfig}
\end{figure}
(A real football does not have constant
spherical curvature outside its vertices; sadly, the geometry 
considered here cannot be isometrically embedded in three flat
spatial dimensions.)  
The size of the wedge is given by the deficit angle $\delta$,
defined by
\be
  \alpha = 1- {\delta \over 2\pi}\ ,
\ee
or
\be
  \delta = {\sigma \over M_6^2}\ .
\ee
It is straightforward to show that this
two-dimensional geometry can be
expressed in spherical coordinates as
\be
    \gamma_{ij} dy^i dy^j = a_0^2(d\theta^2 + \alpha^2
    \sin^2\theta\, d\phi^2) \ ,
    \label{football2}
\ee
where $\phi$ ranges from $0$ to $2\pi$, or equivalently as
\be
    \gamma_{ij} dy^i dy^j = a_0^2(d\theta^2 +
    \sin^2\theta\, d{\tilde\phi}^2)\ ,
    \label{football3}
\ee
where $\tilde\phi=\alpha\phi$ ranges from $0$ to $2\pi\alpha = 2\pi-\delta$.
The fact that the gravitational field of the branes is represented
by a deficit angle, and does not otherwise perturb the surrounding
geometry, can be traced to the fact that there are precisely two
extra dimensions; in higher dimensions it would be much more 
difficult to find an exact solution.

This spacetime has a number of interesting features.  It is, for one thing,
an exact solution to the coupled Einstein/Maxwell/brane equations,
and therefore an interesting starting point for phenomenological
studies.  It would be useful, for example, to understand the
impact of the football geometry on graviton emission at colliders
\cite{particle} or on tests of Newton's law at small scales
\cite{exp}.  Furthermore, although we have not demonstrated it
explicitly, there is every reason to believe that this solution
is stable.  In codimension two there is no force between
point particles or their Lorentz-invariant brane extensions; in
higher dimensions, the branes would attract.  Because the geometry
they are perturbing is stable to begin with, it seems likely that the
brane configuration is also stable.

The most remarkable feature, however, is the independence of the
four-dimensional geometry on the brane tension $\sigma$.
From (\ref{e2}) and (\ref{tune3}), we see that $\sigma$ only enters
into the deficit angle of the football, not into the value of
the magnetic field $B_0$ or the radius $a_0$ necessary to obtain
a flat geometry on the brane.  Indeed, if the brane tension were
to suddenly change, the only effect would be on the internal
geometry of the extra dimensions, not on the macroscopic
(3+1)-dimensional world.  Because the brane tension represents what an
observer on the brane would calculate as the vacuum energy, this
scenario moves the cosmological constant problem completely into
the extra dimensions.  (Where it is still, of course, a problem,
since the magnetic flux has to be finely tuned.)  In this sense
this picture is like the self-tuning models considered in
\cite{self1,self2,self3}.  There are significant differences, however.
The self-tuning models invoke a single extra dimension, with a 
bulk scalar field coupled to matter on the brane, while our model
invokes two extra dimensions with a bulk magnetic field uncoupled to
the branes.  The self-tuning models, furthermore, require a naked
singularity parallel to the brane, while the football geometry is
non-singular (except for the branes themselves, which would be
smeared out in a more realistic treatment).  Perhaps most
importantly, the model considered here was not specifically chosen
to have anything interesting to say about the cosmological constant
problem; we simply examined the simplest possible configuration with
explicit brane sources in stabilized extra dimensions, and the
independence of the four-dimensional geometry on the brane tension
appeared as an unexpected bonus.

Of course we have not solved the cosmological constant problem,
as there is still the need to tune the bulk magnetic field against
the bulk cosmological constant.  In particular, our solution is not
an exception to Weinberg's no-go theorem \cite{weinberg}, because of
the need for this tuning.  However, the cosmological constant 
problem is sufficiently difficult that transforming it into a different
problem is a worthy endeavor, since the new problem might suggest new
solutions.  One could imagine, for example, that a symmetry
(such as supersymmetry) which is absent or broken on the brane is
preserved in the bulk, and is responsible for the apparent fine-tuning.
Continuing in this optimistic vein, one could even imagine that the
branes contribute a small symmetry-breaking effect, disturbing the
balance in the bulk and giving rise to a small effective cosmological
constant of the type observed in current data \cite{carroll}.
This pleasant fantasy is abetted by the numerological coincidence
between the observed vacuum energy, $\rho_{\rm vac}\sim
(10^{-3}{\rm ~eV})^4$, and the maximum allowed size of the extra
dimensions, $R \sim 0.2{\rm ~mm} \sim (10^{-3}{\rm ~eV})^{-1}$.  
Unfortunately, we have no concrete proposal to implement such a scheme.

The way in which the extra-dimensional manifold compensates for
changes in the brane tension is worth remarking on.  In
\cite{sundrum}, Sundrum considered compactifications with spherical
topology and explicit brane sources but no bulk magnetic field or
cosmological constant.  (He also considered bulk fields, but without
explicit brane sources, and did not look for exact solutions with
all the sources at once.)  In that case the problem is equivalent
to looking for closed universes in (2+1) dimensions with no cosmological
constant \cite{djh}.  The Gauss-Bonnet theorem states that a compact,
orientable, even-dimensional genus-0 
manifold with deficit angles $\delta_n$ at conical singularities
and smooth Ricci curvature $R^{(\gamma)}$ elsewhere obeys
\be
  {1\over 2}\int R^{(\gamma)}\sqrt{\gamma}\,d^2y + \sum_n \delta_n
  = 4\pi\ .
\ee
With no bulk fields, the extra dimensions are flat between the
branes ($R^{(\gamma)}=0$), and the total deficit angles must add up to 
$4\pi$, which seems like a fine-tuning.
For the football geometry, however, there is a bulk curvature
\be
  R^{(\gamma)} = {2 \over a_0^2}\ .
\ee
Using the form (\ref{football2}) for $\gamma_{ij}$, we have
\be
  {1\over 2}\int R^{(\gamma)}\sqrt{\gamma}\,d^2y =
  \alpha \int \sin\theta\, d\theta\, d\phi = 4\pi\alpha\ ,
\ee
so that
\be
  4\pi\alpha + \sum_n \delta_n
  = 4\pi\ .
\ee
Since we have two branes, each with deficit angle $\delta = 
2\pi(1-\alpha)$, this relation is automatically satisfied.  In other
words, the deficit angles contribute 
to the integrated curvature themselves,
but they also remove volume from the curved football in an exactly
compensating fashion.  There is thus no need for any fine-tuning
of the brane tensions; the geometry of the extra dimensions
adjusts appropriately.  (We do require that the two branes have
equal tensions, in order that we obtain a static solution; 
if we wanted to, we could ensure this condition by imposing 
reflection symmetry about the equator.)

By considering solutions with Lorentz invariance and
no matter fields on the brane, we have avoided the question of
cosmological evolution.  Since the tension comes from 
vacuum energy of the brane fields, but does not induce
a cosmological expansion on the brane, the Friedmann equation 
$H^2 \sim \rho$ of
conventional four-dimensional cosmology must somehow be modified.  In 
a study of the analogous problem in self-tuning scenarios \cite{cm},
Carroll and Mersini showed that the expansion rate was governed by
the combination $\rho+p$, where $p$ is the pressure, rather than
simply $\rho$ as in ordinary Friedmann cosmology.  Since vacuum
energy has $p_{\rm vac}=-\rho_{\rm vac}$, the cosmological constant
drops out of such a relation, while a matter-dominated
universe ($p_{\rm M}=0$) would evolve normally.  This type of 
cosmology could have difficulty reproducing the successes
of Big-Bang nucleosynthesis; it would be interesting to know what
the analogous results are for the situation considered here.

We have presented an exact solution for a six-dimensional spacetime
with stabilized football-shaped extra dimensions and explicit
brane sources.  In addition to providing an arena for phenomenological
studies, our solution presents an interesting twist to the 
cosmological constant problem, in which the effects of vacuum
energy on the brane only show up in the bulk geometry.  Further work
will be required to determine whether the cosmological evolution of
this sort of spacetime is consistent with observation, and whether
a mechanism can be found to properly adjust the bulk fields to
keep the effective cosmological constant small.

\section*{Acknowledgments}
We thank Eugene Lim, Lisa Randall, Savdeep Sethi, and Kendrick
Smith for helpful conversations.  
This work was supported in part by the NSF grant PHY-0114422 (CfCP)
and DOE grant DE-FG02-90ER-40560 to
the University of Chicago,  the Alfred
P. Sloan Foundation, and the David and Lucile Packard Foundation.

\end{document}